\documentclass[preprint,showpacs,nofootinbib,prd,aps]{revtex4-1}
\usepackage[utf8]{inputenc}
\usepackage{graphicx,amstext,amssymb,wrapfig}

\begin{document}
\title{
Photon spectra and anisotropic flow in heavy ion collisions at the top RHIC energy within the integrated hydrokinetic model with photon hadronization emission}

\author{V.~Yu. Naboka$^1$}
\author{Yu.~M.~Sinyukov$^{1}$}
\author{G.~M.~Zinovjev$^1$}
\affiliation{$^1$Bogolyubov Institute for Theoretical Physics,
03680 Kiev,  Ukraine}

\begin{abstract}
     The integrated HydroKinetic Model (iHKM) is applied to analyze the results of direct photon spectra as well as elliptic and triangular flow measurements in 200A GeV Au+Au collisions at RHIC for different centrality bins. Experiments detect the strong centrality dependence of photon elliptic and triangular flow as increasing $v_n(p_T)$-coefficients towards peripheral collisions. The photon production in the model is accumulated from the different sources along with the process of relativistic heavy ion collision developing. Those include the primary hard photons from the parton collisions at the very early stage of the process, the photons generated at the pre-thermal phase of dense matter evolution, then thermal photons at partially equilibrated hydrodynamic quark-gluon stage, together with radiation displaying a confinement and, finally, from the hadron gas phase. Along the way a hadronic medium evolution is treated in two distinct, in a sense opposite, approaches: chemically equilibrium and chemically non-equilibrium, namely, chemically frozen expansion.  We find the description of direct photon spectra, elliptic and triangular flow are significantly improved, similar to that found in iHKM for the LHC energies, if an additional portion of photon radiation associated with the confinement processes, the ``hadronization photons'', is included into consideration.

\end{abstract}

\pacs{13.85.Hd, 25.75.Gz}
\maketitle

\section{Introduction}
Electromagnetic radiation from the hadronic system produced in relativistic nucleus-nucleus collisions is an unique messenger \cite{Fein, Shur, Ru} while probing new state of produced matter. Nowadays it is recognized that the photon spectra provide us with an information on the state of produced system just at the moment of photon radiation and, hence, they can test some QCD calculations and confinement features \cite{GNSZ, Kharz}. At the same time it is necessary to take into account that relativistic collective flows of expanding superdense matter have a significant influence on photon spectra. As a routine, such flows are described by hydrodynamic-based models absorbing various stages of this matter evolution. Extensive theoretical study of the photon production in such models have been inspired by the unexpectedly large direct photon yield as well as their elliptic flow measured by PHENIX Collaboration at RHIC in the recent years \cite{RHIC_photon}. These results stimulated a detailed analysis of direct photon sources in order to resolve "direct photon flow  puzzle" \cite{Renk, Sriva, Paquet, Kim}.

    In our recent paper \cite{ihkm-photon} we analyzed  the "photon puzzle" observed by the ALICE Collaboration at CERN \cite{ALICE_spectra} within the integrated HydroKinetic Model (iHKM) \cite{iHKM}. The model includes formation of the initial state combining Monte-Carlo Glauber and Color Glass Condensate initial conditions, a gradual (partial) thermalization of the matter created, subsequent viscous hydrodynamic relativistic expansion of quark-gluon plasma, hadronization process, particlization of the  hadron matter and, finally, scattering  of secondary  particles and spectra formation. The iHKM describes pretty well the bulk hadron observables at different centralities. It concerns various particle yields in the form of number ratios, pion, kaon, (anti)proton spectra, charged hadron elliptic flows, pion and kaon interferometry radii at the top RHIC energy 200A GeV  Au+Au collisions and available LHC  energies for Pb+Pb collisions 2.76 TeV and 5.02 TeV \cite{ihkm-rhic,ihkm-lhc276, ihkm-lhc502}. The  predictions of the model, e.g.  \cite{kaon-femto, Kstar}, were experimentally confirmed later in the experiments, e.g. \cite{ALICE-kaon}. The iHKM parameters  adjusted for baryon observables are of the same value in all simulations,  except for the initial maximal energy density and baryon chemical potentials which depend on the collision energy at RHIC and LHC.  Using the corresponding parameters of Ref. \cite{ihkm-photon} the iHKM has been applied for describing the direct photon transverse spectra and their anisotropy, expressed through the so-called $v_2$-coefficients, at the LHC energy $\sqrt{s_{NN}}= 2.76$ TeV \cite{chliap}.

                The direct photon production accumulates the processes of quark-quark and quark-gluon collisions at the very early stage of hard parton collisions resulting in the so-called prompt photons, then photons produced at the pre-thermal  evolution stage of the created matter, thermal photons from QGP,  and finally  photons from the last stage - hadronic evolution. It was found that the data description, both the photon spectra and $v_2$, can be improved significantly if one takes into account an additional photon radiation originated by the hadronization process that transforms the hypothetical mechanisms into an illumination of fundamental theory problems. Clearly, the nature of photon radiation at hadronization is far to be persuasively elaborated and there are several phenomenological scenarios under discussion which are quite efficient \cite{azimuth-aniz, Kharz, Camp-1, Pratt, Itakura}. We use such an idea trying to give more credibility to these mechanisms. We speculate here adapting the phenomenological  prescription for describing the photon emission from the hadronization space-time layer in the cross-over scenario at the top RHIC and LHC energies \cite{ihkm-photon}. Since a matter of a space-time layer, where a hadronization takes place, is actively involved in anisotropic transverse flow, both positive contributions to the spectra and $v_2,~v_3$ coefficients are considerable albeit such a way needs a further research and elaboration.

The paper is organized as follows. Section \ref{sec:iHKM} is devoted to the brief review of iHKM in its application for modeling the matter evolution.  The description of  different sources of the direct photon radiation associated with the corresponding stages of the evolution is given in Section \ref{sec:sources}. The results and discussion are presented in Section \ref{sec:results}. Section \ref{sec:summary} concludes the paper.

\section{Integrated hydrokinetic model}
\label{sec:iHKM}
Before addressing the possible sources of photon emission, we will describe briefly the model of matter evolution used in this research. We utilize the integrated hydrokinetic model (iHKM) \cite{iHKM} for modelling the process of heavy ion collision. This model was first developed and used for highest energies available at Large Hadron Collider, but it is also applicable for top RHIC energy \cite{ihkm-rhic}. As it was already mentioned, iHKM considers heavy ion collision as consisting, at least, of five stages: initial stage generation, prethermal stage, thermal (hydrodynamic) stage, particlization stage and UrQMD hadronic cascade stage. Further we describe each stage more specifically.

\subsection{Initial state generation}

For initial state generation we use GLISSANDO \cite{gliss,gliss2,gliss3} package, which works in the frame of semiclassical Monte-Carlo Glauber model. The initial state, generated by GLISSANDO package, is then attributed to the initial time $\tau_0=0.1$ fm/c  \cite{iHKM, ihkm-rhic}. Note, the value of the free parameter $\tau_0$ has to be larger than the time $\tau_{ol}$ when the overlapping of colliding nucleus happens. Specifically, $\tau_{ol} \approx 0.07, ~0.005, ~0.003$ fm/c for energies RHIC $\sqrt{s_{NN}}=200$ GeV, LHC $\sqrt{s_{NN}}=2.76$ TeV and $\sqrt{s_{NN}}=5.02$ TeV correspondingly. Therefore, an utilization of the initial energy density profile as an already formed superdense medium at the time $\tau_0=0.1$ fm/c may be plausible enough for all the mentioned energies, and we use the same value 0.1 fm/c when apply iHKM  to the top RHIC and LHC energies \cite{preth, iHKM, ihkm-rhic}. Of course, a scale of the initial energy density profile depends on the collision energy.

The initial state at $\tau_0$ is defined by the combination of several factors \cite{iHKM}. They are: 1) the coefficient $\alpha$, which defines a relative contribution of binary collisions to the formation of the initial energy density profile (which also includes contribution from wounded nucleons), 2)  the maximal energy density $\epsilon_0$ at the initial time $\tau_0$ at most central collisions, 3) the coefficient $\Lambda$, which defines the momentum spectrum anisotropy of partons at the initial moment $\tau_0$ according to the Color Glass Condensate (CGC) model \cite{iHKM,ihkm-rhic}. In this paper we use $\alpha=0.18$ and $\epsilon_0 = 235$ GeV/fm$^3$ as these values provide the good description of multiplicity dependence on collision centrality for the top RHIC energy for Au+Au collisions \cite{ihkm-rhic}. We set $\Lambda=100$, as in our previous publications \cite{preth, iHKM, ihkm-rhic}.

\subsection{Relaxation model of prethermal stage}

The relaxation model is describing a thermalization of an initial state \cite{preth, iHKM, relaxation}. It maps the continuous transition from locally non-equilibrated initial state at moment $\tau_0$ to the near equilibrated state at the moment $\tau_{th} \gg \tau_0$.
The results for the top RHIC and LHC energies are not sensitive to the value of $\tau_{th}$, if along with $\tau_{th}$ one changes correspondingly another parameter – maximal energy density at the initial time $\tau_0$ \cite{ihkm-lhc502, IC}. Here, as in the other papers based on iHKM, the value of $\tau_{th}$ is chosen to be equal 1 fm/c.

The energy-momentum tensor of matter on the prethermal stage can be written as \cite{iHKM, relaxation}
\begin{eqnarray}
T^{\mu \nu}(x)=T^{\mu \nu}_{\text{free}}(x){\cal
P}(\tau)+T_{\text{hydro}}^{\mu \nu}(x)[1-{\cal P}(\tau)], \label{tensor}
\end{eqnarray}
where $T^{\mu \nu}_{\text{free}}(x)$ is the free-evolving part of the total energy-momentum tensor and $T^{\mu \nu}_{\text{hydro}}(x)$ is hydrodynamically evolving component. The function $\cal P(\tau)$ has the form \cite{preth,relaxation}
\begin{eqnarray}
{\cal P}(\tau)=  \left(
\frac{\tau_{\text{th}}-\tau}{\tau_{\text{th}}-\tau_0}\right
)^{\frac{\tau_{\text{th}}-\tau_0}
 {\tau_{\text{rel}} }}.
 \label{P}
\end{eqnarray}
As one can see, $P(\tau_0)=1$ and $P(\tau_{th})=0$. Thus, the matter is gradually transiting from the pure free-streaming in the initial time $\tau_0$ to pure hydrodynamic evolution at the thermalization time $\tau_{th}$. Parameter $\tau_{rel}$ describes the rate of this transition. As for the LHC energy, for RHIC we put $\tau_{rel}=0.25$ fm/c \cite{iHKM, ihkm-rhic}. Writing down the conservation laws for the total energy-momentum tensor and accounting for the properties of the free-evolving energy-momentum tensor, we have
\begin{eqnarray}
\partial_{;\mu}\widetilde{T}^{\mu
\nu}_{\text{hydro}}(x)= - T^{\mu \nu}_{\text{free}}(x)\partial_{;\mu}
{\cal P}(\tau), \label{pre-equation}
\end{eqnarray}
where $\widetilde{T}^{\mu\nu}_{\text{hydro}}(x)=[1-{\cal P}(\tau)]T^{\mu\nu}_{\text{hydro}}(x)$. The relaxation model in the form used in this paper calculations, also contains a shear viscosity tensor terms in the Israel-Stewart form. Evolution of the shear stress tensor can be derived similarly to (\ref{pre-equation}):
\begin{eqnarray}
[1-{\cal P}(\tau)]\left \langle u^\gamma \partial_{;\gamma}
\frac{\widetilde{\pi}^{\mu\nu}}{(1-{\cal P}(\tau))}\right \rangle
=-\frac{\widetilde{\pi}^{\mu\nu}-[1-{\cal
P}(\tau)]\pi_\text{NS}^{\mu\nu}}{\tau_\pi}-\frac {4}{3}
\widetilde{\pi}^{\mu\nu}\partial_{;\gamma}u^\gamma.
\label{pre-viscous}
\end{eqnarray}

If $\tau \rightarrow \tau_{th}$ then ${\cal P} \rightarrow 0$ and system reaches  the  target energy-momentum tensor of viscous relativistic hydrodynamics in the  Israel-Stewart form. The Laine-Schroder equation of state (EoS) \cite{Laine} is used for hydrodynamic component at prethermal stage and consequent thermal hydrodynamic stages.

\subsection{Hydrodynamic stage}

The stage of near locally equilibrated hydrodynamic evolution follows the prethermal stage, it starts at $\tau=\tau_{th}$ and lasts till the hypersurface of constant temperature $T=165$ MeV. By setting the source terms in (\ref{pre-equation}) and (\ref{pre-viscous}) to be zero, we get the casual viscous hydrodynamic evolution equations, see previous subsection. For viscosity coefficient on this and previous stage we use its near minimal value, $\eta/s=0.08$ \cite{iHKM}. The difference from the LHC case is the necessity to  use for top RHIC energy the different chemical potentials into consideration. We settle chemical potentials equal to be the same as in Ref. \cite{ihkm-rhic}  and use the same prescription to account them as in that paper.
It is worthy to note that the change of EoS to another one, e.g. EoS from Hot QCD Collaboration \cite{EoS2} does not destroy the results, if simultaneously one provides an appropriate adjustment of the initial time for the superdense matter formation and related maximal initial energy density \cite{ihkm-lhc502}.

\subsection{Particlization}

As far as hydrodynamic model is not applicable when matter becomes diluted, we have to develop a transition from hydrodynamic stage to further evolution of matter as hadron-resonance gas. We make this transition at a hypersurface of constant temperature, specifically, $T=165$ MeV. (It corresponds to energy density $\epsilon = 0.5$ GeV/fm$^3$ at the Laine-Schroder EoS). We suppose that this particlization temperature is close to effective hadronization temperature, some kind of approximation in cross-over scenario.  To build the switching hypersurface we utilize the Cornelius routine \cite{convert}. For conversion of fluid matter to hadron gas, the Cooper-Frye formula, and  Grad's 14-momentum ansatz for viscous corrections are used.

\subsection{Hadron gas stage}

In original full iHKM version the further evolution of matter as hadron gas is simulated by using Ultrarelativistic Quantum Molecular Dynamics (UrQMD) \cite{urqmd}. Note, that in the iHKM  calculations of hadron spectra we did not use the conception of freeze-out, neither “thermal”, nor “chemical”. The particles are leaving the system gradually from the 4D space-time region and inelastic scattering continues also below the effective hadronization/particlization temperature 165 MeV. The calculations in iHKM  in each event are carried out up to a large time, in fact, till 500 fm/c but the intensive hadronic cascade lasts 4-5 fm/c after hadronization \cite{Kstar}. The picture of continuous particle emission brings good agreement with experimental data for hadrons, also at RHIC \cite{ihkm-rhic}. In this article the same set of parameters as in the paper \cite{ihkm-rhic} is used, however, we are forced to calculate the gradual emission of {\it photons} at this final stage rather approximately.  The reason is that UrQMD is not appropriate for describing the photon emission. To solve this problem and estimate an inaccuracy  we consider the photon radiation at the {\it hadronic} stage with continuous particle emission in the  two different approaches, similar as it is done in  Ref. \cite{ihkm-photon}.

		The first one is rather simple -- we just prolongate calculations of the chemically equilibrated evolution until the time when photon radiation become neglectible, say, until $T=100$ MeV (again, it is not freeze-out temperature neither for hadrons nor for photons!). We also compare the results with the analogical results for calculations with temperatures 115 and 80 MeV, the final thermal photon production at this last stage does not change significantly. Albeit this approach seems to be rather primitive, it provides us with an approximate estimate of photon emission and anisotropy.

 The second approach, on the other hand, is to use the original hydrokinetic model (HKM) \cite{HKM 2008, original HKM, HKM} that describes a continuous hadron emission  below hadronization temperature $T=165$  MeV, and provides chemically non-equilibrated  hydrodynamic evolution in the form which is opposite to chemically equilibrated, namely, chemically frozen evolution. The latter excludes changing the particle numbers for each particle species  except contributions that bring resonance decay processes (see all details in \cite{original HKM, HKM}). Such approach is built  in full accordance with the chemical freeze-out conceptions. In HKM formalism this leads to appearance of individual chemical potentials for hadrons during the matter evolution in addition to baryon and electric chemical potentials, that are induced by the conservation laws.  It changes the concentrations of the hadrons that participate in reactions where photons are produced, as compared to their concentrations at the chemically equilibrated scenario. Below we describe this in more detail. Note that chemically non-equilibrium hydrodynamic evolution was proposed already in 2002 in Ref. \cite{partial CHE}, where it was considered as an intermediate scenario - {\it partial} chemically equilibrium evolution, that excludes some inelastic reactions.

The HKM allows to account {\it local non-equilibrium}  effects in hadron spectra formation using  escape probabilities formalism within the Boltzmann equation in integral form \cite{HKM 2008}. The calculations of escaped probability values utilize the ideal hydrodynamics as a first approximation. In the same order of approximation photons have unite escape probabilities because they penetrate the hadron medium, and our results are obtained just for this case. Photon spectra at locally chemically equilibrium evolution are described by the formula (\ref{spectra2}). The second order of approximation for hydrodynamic velocity $u$ will appear when one takes into account the back reaction of the potentially escaped particles to hydrodynamic evolution and so making this background evolution locally non-equilibrated. This is technically quite complicated problem and was not applied in HKM.
The HKM model is the predecessor of iHKM and was developed for the ideal hydro only. That is why we use $\eta$/s=0 in this case.
In this article we consider the two opposite scenarios: chemically equilibrium and chemically frozen evolution to estimate uncertainties  for photon production at the very last stage - hadron gas expansion.

\section{Sources of direct photon emission}
\label{sec:sources}
There are many different sources of direct photons emitted during the heavy ion collision, each of which has the maximal impact during a certain stage of the collision process. Overall, they can be subdivided into prompt photons which are emitted in the earliest stage of collision, even before very dense matter forms, photons that are emitted at pre-thermal stage of the matter evolution,  thermal photons  emitted at the hydrodynamic stage of quark-gluon plasma  and hadron gas expansion due to parton-parton and hadron-hadron scattering and annihilation, and, finally, photons emitted due to hadronization mechanism discussed in our previous paper  \cite{ihkm-photon}. Further we consider these photon sources in more detail.

\subsection{Prompt photons}

\begin{wrapfigure}{r}{0.5\textwidth}
     \includegraphics[width=0.5\textwidth]{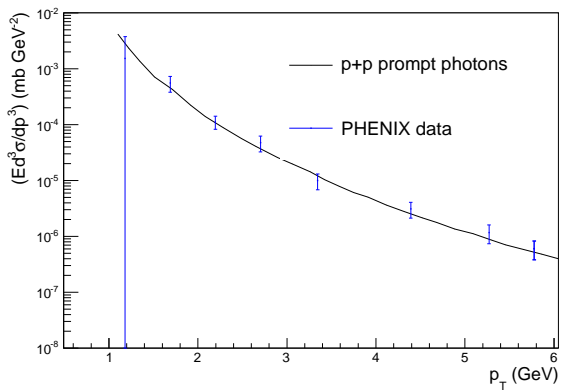}
     \caption{\small Total photon spectra for proton-proton collision in comparison with experiment. Experimental results are taken from \cite{PHENIX-pp}.}
     \label{fig:prompt-single}
\end{wrapfigure}

Prompt photons are the photons resulting from the very first instances of the nuclear collision, mainly through the Compton scatterings of partons and quark-antiquark annihilations. These photons can be calculated using the perturbative QCD (pQCD) techniques. Accumulating the results of various experiments \cite{ATLAS_photon, CMS_photon}, it could be claimed that prompt photon spectra scale with the binary nucleon-nucleon collision number \cite{Ncoll}. This value is calculated  with the Monte Carlo Glauber code \cite{gliss, gliss2}, see Subsec. II A. Thus, prompt photon spectra in Au+Au collisions can be presented as convolution of binary collision number $N_{coll}$ with the proton-proton spectra calculated within pQCD and dealing with the calculations in the form of cross section

\begin{equation}
d\sigma = \sum_{i,j,k}f_i\otimes f_j \otimes d\hat{\sigma}(ij\rightarrow k)\otimes D^{\gamma}_k,
\label{g_s}
\end{equation}
where the summation runs over all possible partonic subprocesses, $f_i$ and $f_j$ are the parton distribution functions, $D^{\gamma}_k$ is the fragmentation function and $d\hat{\sigma}$ ~---
cross section of the corresponding partonic subprocess. The cross section in Eq. (\ref{g_s}) is calculated by perturbative expansion in the strong coupling constant. The QCD scales, which are present in such an expansion, are set to $Q_{fact}=Q_{ren}=Q_{frag}=0.5~p_T$. In order to compute the cross section (\ref{g_s}) for momentum range $0.5$ GeV/c $< p_T < 6$ GeV/c, we address the JETPHOX package \cite{JETPHOX}. The pQCD calculations have rather high lower $p_T$-limit, especially for such low proportionality coefficient in coupling constant/momentum dependence as $Q=0.5~p_T$, which we have chosen in our calculations. In fact,we realize the calculations can be done for higher proportionality coefficient, specifically $Q=8.0~p_T$, and then just rescale the obtained spectra (see  the details in Ref. \cite{Paquet}). In our calculations we utilize EPS09 parton distribution functions \cite{EPS09} and BFG II fragmentation functions \cite{BFG-2} (both are represented as the tables of values). The results of such calculations for proton-proton collisions are shown on Fig. \ref{fig:prompt-single}

\subsection{Prethermal and thermal photons}

As far as iHKM contains prethermal stage, it is natural to consider this stage also providing some photon emission but this process should be treated differently from the photons originated by the thermalized QGP. These photons should not be mistakenly considered as the prompt photons generated in the first instances of collision. Prethermal photons in our approach are emitted by the matter in time interval $0.1$ fm/c $< \tau < 1.0$ fm/c. The total energy-momentum tensor for the prethermal stage of iHKM consists of hydrodynamic and free-streaming components (\ref{tensor}). Actually we consider prethermal photons as thermal that are coming from the hydrodynamic component at the pre-thermal stage. Then the resulting spectra is multiplied by $1-{\cal P(\tau)}$, as this factor is a weight coefficient of hydrodynamic component in the total energy-momentum tensor. The hydrodynamic component has the same equation of state as in chemically equilibrated quark gluon plasma. The temperature is calculated according to it from the energy density in the evolutionary equations.

The spectra of photons originated from QGP were successfully described  \cite{Arnold} in the leading order of strong coupling constant $g_s$ and the corresponding formulas are used in the calculations of this paper. The main sources of such photon emission are of leading order 2 $\rightarrow$ 2 processes in the hot QGP. It should be also noted that next to leading order processes can contribute the same order terms to the thermal photon rate \cite{Gelis, Ghiglieri}. To describe the spectra and anisotropic flow of QGP-originating thermal photons one needs a reliable tool for modelling the matter evolution. As such, we utilize the above mentioned iHKM (Sec. \ref{sec:iHKM}). The photon emission in the relativistic-invariant form can be described by the formula

\begin{equation}
k^0\frac{d^7N}{d^3\textbf{k}d^4\textbf{x}}={k}\cdot {u}~\frac{\nu_e(k\cdot {u})}{(2\pi)^3},
\label{spectra2}
\end{equation}

where $\textbf{k}$ is the photon momentum and $\textbf{u}$ is the collective flow of matter in the coordinate $\textbf{x}$. Thus, $\nu_e(k\cdot {u})$ has a physical meaning of spontaneous emission rate of photons with momentum $\textbf{k}$. The evaluation of $\nu_e(k\cdot {u})$ was made in \cite{Arnold} by summarizing emission output from partonic 2 $\rightarrow$ 2 subprocesses with inclusion of near-collinear bremsstrahlung and inelastic pair annihilation contributions, and the Landau-Pomeranchuk-Migdal suppression effects are also included. The final spectra then has the form
\begin{equation}
\frac{dN}{2\pi k_Tdk_T}=\sum_i\frac{1}{(2\pi)^4} \Delta^4 V(x_i)\int_0^{2\pi} {k}\cdot {u(x_i})\nu_e({k}\cdot {u(x_i}))d\phi,
\label{spectra_qgp}
\end{equation}
$\Delta^4 V=\tau \Delta \tau\Delta x\Delta y\Delta\eta$ is a volume of a cell in the computation grid. The summation runs over a 3-dimensional computation grid. As far as this formula is applicable for QGP, it runs over all cells with $T>165$ MeV, thus affecting prethermal and hydrodynamic stages of iHKM (\ref{sec:iHKM}). Note, that the rate (\ref{spectra_qgp}) does not account for  viscous corrections to parton disctribution function in QGP since we consider hydrodynamics with minimal ratio $\eta/s$. Therefore we do not expect that such corrections in our scenario can solve the ``photon puzzle'' and just neglect them.

The emission from cells with lower temperature  $T<165$ MeV is described by the same formula (\ref{spectra_qgp}), but with $\nu_e(k\cdot {u})$ derived for hadron gas emission. The hadron gas photon emission rate consists of many terms, which have a different nature. These include:\\

1) emission from a meson gas \cite{Turbide}. More precisely, we consider reactions
$\pi+\rho \rightarrow \pi+\gamma$, $\pi+\pi \rightarrow \rho+\gamma$, $\rho \rightarrow \pi+\pi+\gamma$, $\pi+K^{*} \rightarrow K+\gamma$, $\pi+K \rightarrow K^{*}+\gamma$, $\rho+K \rightarrow K+\gamma$, $K+K^{*}\rightarrow \pi+\gamma$. Each of these reactions gives its contribution to the total photon emission, with a rate which is described in \cite{Turbide}. In a case of chemically frozen scenario, we additionally multiply emission rate of each reaction by a factor $exp\left(\frac{\sum \mu_i(\tau,x)}{T(\tau,x)}\right)$, where $\mu_i(\tau,x)$ are chemical potentials of the reaction constituents. For example, emission rate of the reaction $\pi+\rho \rightarrow \pi+\gamma$ is multiplied by $exp\left(\frac{\mu_\pi(\tau,x)+\mu_\rho(\tau,x)}{T(\tau,x)}\right)$. The typical behavior of the chemical potentials on the energy density in HKM is presented in Fig. 8 in Ref. \cite{HKM 2008}.\\

2) contribution from in-medium $\rho$ mesons \cite{Heffernan}; this specific emission depends on baryon chemical potential $\mu_B(\tau,x)$ in the case of chemically equilibrium hadron matter evolution and, in the case of chemically frozen expansion, also on additional species-dependent chemical potentials of the reaction constituents.\\

3) thermal photon emission from the pi-rho-omega system. This includes reactions $\pi+\rho\rightarrow\omega+\gamma, \rho+\omega\rightarrow\pi+\gamma, \pi+\omega\rightarrow\rho+\gamma$ \cite{Rapp}; In a case of chemically frozen scenario, it is multiplied by a factor analogical to the meson gas case\\

4) photon emission from $\pi\pi$ bremmstrahlung \cite{Heffernan}.

\subsection{Hadronization photons}

The constructed phenomenological models (sometimes very inventive) are inspired by the constantly evolving possibilities of the experiment (starting since an excess of photon production observed in \cite{chliap}), which supports a lively interest in this issue. In our previous work \cite{ihkm-photon} for LHC energies we have developed a prescription for calculating the photon emission on the hadronization stage. This phenomenological approach is based on a suggestion about an additional soft photon radiation from a confining process proposed in Refs.  \cite{azimuth-aniz, Kharz, Camp-1, Pratt, Itakura}.

  One of the first attempts to consider production of photons at the point of hadronization was made in Ref. \cite{Camp-1}. It was supposed as the mechanism where the soft gluon interactions could mediate quark-antiquark annihilation when the system moves toward color neutrality, resulting in a large increase in photon production. In Ref. \cite{Pratt} as opposed to how gluonic degrees of freedom might create light as they disappear, the authors were primarily interested in how the electromagnetically charged quarks and anti-quarks radiate when they hadronize. The photons are produced at the point of recombination because quarks that carry electric charge are accelerated during hadronization. In Ref. \cite{Itakura} authors also discuss photon emission at the stage of hadronization. They formulate  this in accordance with the mechanism  of “radiative recombination,” similar to that is occurring when a plasma goes back to an atomic gas. When a QGP hadronizes,  the analogous processes such as $q + \bar{q} \rightarrow \pi^0 + \gamma$ is expected. The photon emission in that work is essential to compensate the energy difference between the initial and final quark-gluon states. Despite some such models of photon emission at hadronization are able to improve photon spectra significantly, they are still phenomenological and incomplete, since the process is a nonperturbative and nonequilibrium.
	
	In general, the problem of hadronization is complex. It is known that quarks and gluons produced in hard-scattering reactions give rise to the colorless hadronic bound states that form jets. The associated hadronization process is described by fragmentation functions which are universal functions representing, in the simplest picture, a measure of the probability density that an outgoing parton produces a hadron. The fragmentation function is studied in jet production at $e^++e^-$, $e+p$ and $p+p$  collisions. At low values of $p_T$ jet, one needs to consider the non-perturbative models of fragmentation, the connection of the partons to the remainder of the event and to the accretion of particles from the underlying event into the jet. The photon radiation arising during hadronization inside the jet is described by the so far little known {\it photon fragmentation functions}.
	
As for the phase transition/crossover between quark-gluon plasma and hadron gas, there are attempts (e.g. \cite{ Nayak}) to analyze the parton to hadron fragmentation function from quark-gluon plasma using Lattice QCD method at finite temperatures. The photon fragmentation function during these transitions  is not analyzed yet. All available models of hadronization for jet production in $e^++e^-$, $e+p$ and $p+p$  reactions, on the one hand, and in confinement processes during the matter expansion in relativistic A+A collisions and Early Universe, on the other, are not unified at the moment into the full theoretical scheme. Because of this, an analysis based on universal parton–to-hadron and parton-to-photon fragmentation functions is outside the scope of this paper, and we still use the phenomenological approach for RHIC top energy similar as it was done in Ref. \cite{ihkm-photon} for LHC.

Let $G_{hadr}(t,{\bf r},p)$ be the emission function of additional photon radiation at the hadronization stage
\begin{equation}
\frac{d^3N_{\gamma}}{d^3p}=\int dtd^3r~G_{hadr}(t,{\bf r},p)
\label{G_hadr}
\end{equation}
Let  $\sigma$ be the hypersurface of temporal points $t_{\sigma}({\bf r}, p)$ of maximal emission for photons with momentum $p$. We change variables from $({\bf r}, p)$ to $({\bf x}, p)$, where ${\bf x}$ includes the saddle point for given ${\bf r}$ \cite{var-1,var-2}: ${\bf x}={\bf r}+\frac{{\bf p}}{p^0}t_{\sigma}({\bf r},p)$. Then we use the saddle point approximation for emission function $G_{hadr}\approx F(t,{\bf x},p)\exp(-\frac{(t-t_{\sigma})^2}{2D^2})$, where $F$ has smooth dependence on $t$. It leads to
\begin{equation}
\frac{d^3N_{\gamma}}{d^3p}=\int d^3x \left|1-\frac{{\bf p }}{p^0}\frac{\partial t_{\sigma}({\bf x}, p)}{\partial {\bf x}}\right|\int dt F(t_{\sigma}({\bf x}, p), {\bf x}, p) \exp\left(-\frac{(t-t_{\sigma}({\bf x}, p))^2}{2D_c^2(t_{\sigma}({\bf x}, p), {\bf x}, p)}\right)
\label{add}
\end{equation}
Assuming that the hypersurface of maximum photon emission corresponds to the hadronization isotherm ($T_h=165$ MeV in our model), we can write down (\ref{add}) in the invariant form:
\begin{equation}
p^0\frac{d^3N_{\gamma}}{d^3p}=\int_{\sigma_h}d^3\sigma_{\mu}(x)p^{\mu}F\left(p\cdot u(x),T_h\right)D_c\left(p\cdot u(x),T_h\right)\theta(d\sigma_{\mu}(x)p^{\mu}),
\label{cooper_frye}
\end{equation}
where $\theta(z)$ is the Heaviside step function that excludes negative contribution to spectra from non-space-like parts of hadronization hypersurface.

Note that expressions (\ref{G_hadr}) – (\ref{cooper_frye}) are quite general for the particle emission from fairly narrow (in time-like direction)  $4D$-area surrounding the hypersurface of the maximal emission of the particles.
We try to include  the hadronization emission mechanism in the simplest phenomenological form and suppose, as proposed in Ref. \cite{Itakura} and our paper \cite{ihkm-photon}, that $FD$ function in (\ref{cooper_frye}) has thermal-like form
\begin{equation}
F D_c = d_c\gamma_{hadr}~f_{\gamma}^{eq}\left(p\cdot u(x),T_h\right)=\gamma_{hadr}d_c\frac{1}{(2\pi)^3}\frac{g}{\exp\left(p\cdot u(x)/T_h\right)-1},
\label{surface}
\end{equation}
where $p\cdot u \equiv p^{\mu}u_{\mu}$, $g=2$, $T_h = 165$ MeV.  Note that irrespective of the thermal-like form of parametrization, we consider the hadronization photon rate separately from thermal production as having another origin.

The value $\gamma_{hadr}$ is defined by the hadronization process, and $d_c \propto \left\langle D\right\rangle$ is defined by temporal width of this process, so it could depend on the collision centrality.
We use the value $\beta\equiv d_c\gamma_{hadr}=0.04$ as providing the best description of PHENIX-RHIC data for the $0-20\%$. We keep this value the same for other centralities.

  Note that in our previous paper \cite{ihkm-photon}, where the effective coefficient $\beta = 0.02$ for the most
	central collisions was used for LHC energy, it was remarked that, with the same coefficient, a description of the spectra and elliptic flows at the LHC in non-central collisions can also be achieved within experimental errors. Therefore, the main point, that required analysis, is the very difference between the coefficients $\beta = 0.02$ for ALICE-LHC and $\beta = 0.04$ at PHENIX-RHIC in our analysis. It is worth noting that the direct  photon  yields presented by STAR Collaboration are in agreement with our calculations (within error bars) for RHIC with coefficient $\beta = 0.02$, the same as for LHC (see Fig. \ref{fig:spectra_all_STAR}). But as we plan to describe simultaneously not only photon yields but also elliptic and triangular flow, presented by PHENIX Collaboration, we use in this article $\beta = 0.04$.  Accounting for the hadronization is long-distance process, involving only small momentum transfers, we can assume that the difference with the LHC parameter $\beta$ is because of difference in maximal intensity (that used in saddle point approximation) of hadronization processes at RHIC and LHC. The intensity, very likely, decreases with increasing of the transverse expansion rate in the system (which is significantly higher at the LHC) because then the coalescence mechanism is hindered. However, we do not undertake to clarify in more detail the interplay between various factors affecting the value of the effective parameter $\beta$ at different collision energies, also because above mentioned STAR vs PHENIX photon data uncertainties.	
		
\begin{figure}
     \centering
     \includegraphics[width=1.0\textwidth]{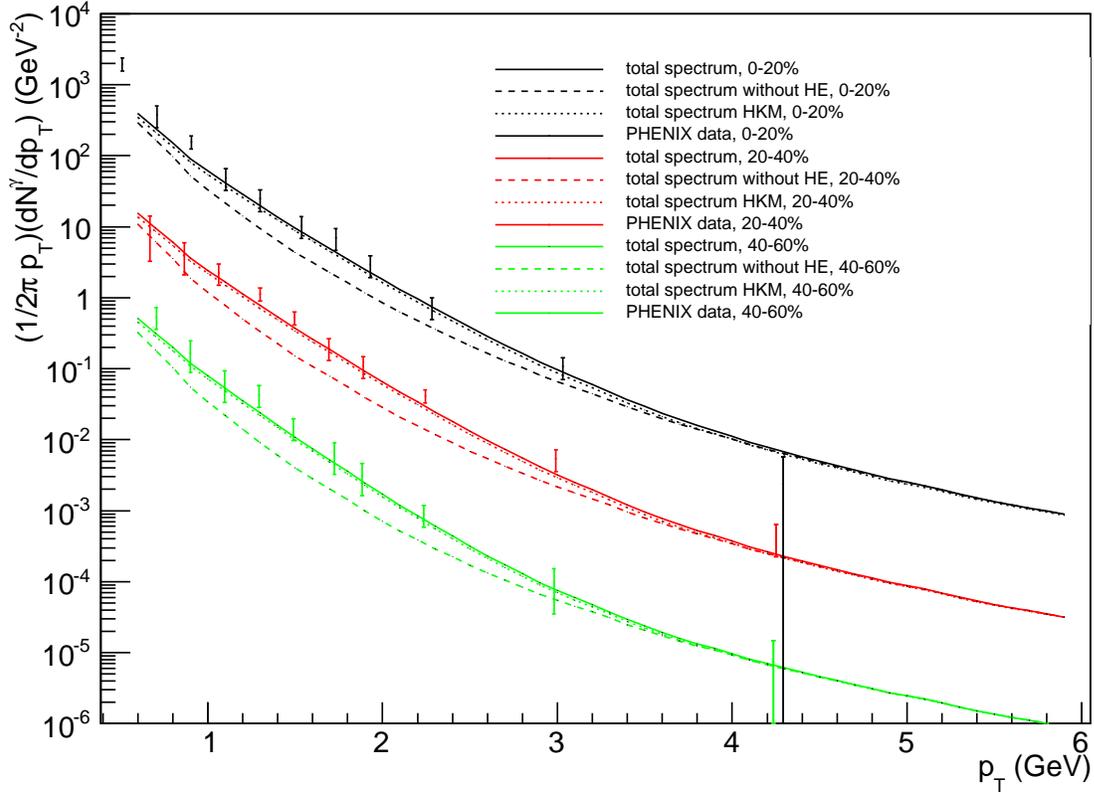}
     \caption{\small
Total photon spectra with the different set of ingredients: iHKM chemically equilibrated with hadronization emission (HE) contribution, iHKM chemically equilibrated without HE contribution, HKM chemically frozen at the hadron stage with HE contribution. Results for centralities 0-20\%, 20-40\% and 40-60\% are included. Spectra for 0-20\% centrality are multiplied by factor of 100, and spectra for 20-40\% centrality are multiplied by factor of 10. Experimental results are taken from \cite{RHIC-spectra}.}
\label{fig:spectra_all}
\end{figure}

\begin{figure}
     \centering
     \includegraphics[width=1.0\textwidth]{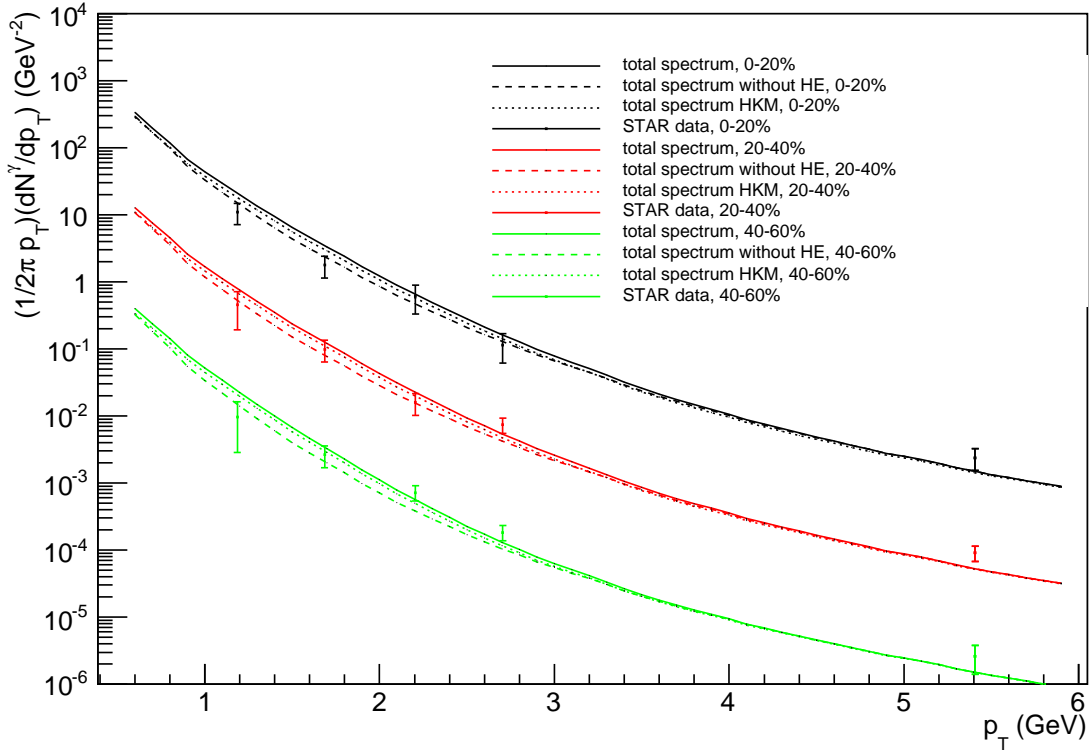}
     \caption{\small
The same as Fig. \ref{fig:spectra_all}, but with $\beta = 0.02$ and comparison to STAR data. Experimental results are taken from \cite{STAR-spectra}}
\label{fig:spectra_all_STAR}
\end{figure}

\begin{figure}
     \centering
     \includegraphics[width=1.0\textwidth]{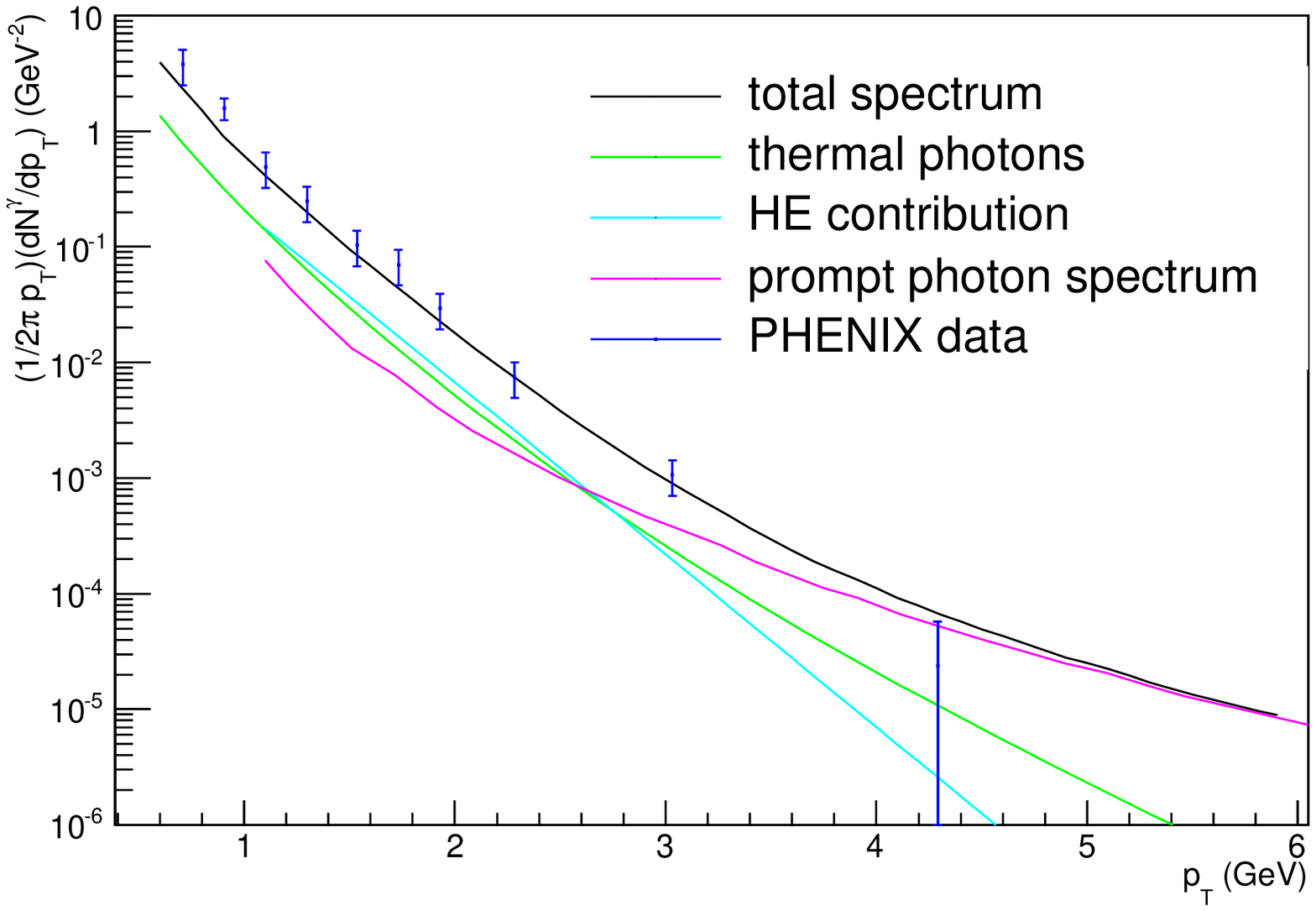}
     \caption{\small
Total photon spectra calculated within chemically equilibrated iHKM for 0-20\% centrality along with its constituents:
thermal (including prethermal) photons, prompt photons, and hadronization emission (HE) contribution. Experimental results are taken from \cite{RHIC-spectra}.}
\label{fig:spectrum_compare}
\end{figure}

\section{Results and discussion}
\label{sec:results}

The total direct photon spectra in Au+Au collisions at $\sqrt{s_{NN}}= 200$ GeV are calculated as a sum of above mentioned contributions, specifically, prompt photon spectra, thermal photon spectra (which includes photons from prethermal stage), and  photons arising due to hadronization process. All the model parameters are described in Sec. II. This set of parameters is chosen because it provides a good description of bulk observables, such as particle yields and number ratios, charged particle multiplicity, pion, kaon, (anti)proton spectra, charged particle momentum anisotropy and pion and kaon HBT-radii at RHIC  for different centralities \cite{ihkm-rhic}. Thus, one of the strongest points of our approach is using the same model with the same set of parameters to describe, besides the photon emission, the  numerous bulk hadron observables as well. As we already discussed in detail in Subsec. II E,  because of the technical reasons, the only approximation made is that in order to estimate the photon emission (and its possible uncertainty) at post-hadronization stage of evolution, the different scenarios of  hadronic  matter hydrodynamic evolution  were used instead of the UrQMD cascade. We compare the two opposite approximation at this point: chemically equilibrated and chemically frozen hydrodynamic expansion. The important ingredient of the full model is presence of photon hadronization emission (HE), as it was discussed in Subsec. III C. The intensity of emission is fixed by the parameter $\beta=0.04$ as providing the best description of $0-20\%$ centrality events. We keep this value  the same also for other centralities at RHIC, as was discussed above.

Anisotropic flow coefficients are defined from a Fourier expansion of the azimuthal correlation function
\begin{eqnarray}
\frac{dN}{d\phi} = \frac{N}{2\pi}\left(1+\sum_n 2v_n \cos(n(\phi-\psi_n))\right),
\label{v2-1}
\end{eqnarray}
The geometric anisotropy in participant plane for n-flow harmonic transforms into the corresponding anisotropic flow, and as the matter of direct calculations, the angle $\psi_n$ in event plane nearly coincides with the corresponding angle of ``minimal axis'' in participant plane \cite{gliss3}:
\begin{eqnarray}
\psi_n = \frac{1}{n}\arctan\frac{\int dx dy \textbf{r}^n \sin(n\phi)\epsilon(\textbf{r},\tau_0)}{\int dx dy \textbf{r}^n \cos(n\phi)\epsilon(\textbf{r},\tau_0)}+\frac{\pi}{n},
\label{v2-2}
\end{eqnarray}
where $\epsilon(\textbf{r},\tau_0)$ is the initial energy density distribution.

In order to avoid too sharp initial conditions, which could lead to some computational artifacts because of large density gradients in viscous hydrodynamics at very small time, we unite events in participant plane in the groups of 50 events and perform a rotation of the initial state of each event by the $\psi_2$ angle for elliptic and $\psi_3$ for triangular flow calculation. Then, for each such ``single'' event we can calculate the photon anisotropy in event plane. Since we do not deal with separate photons, but just with intensities of their radiation, we provide all necessary averaging of corresponding trigonometric functions over the photon multiplicities (at given intervals in $p_T$) in different azimuthal angular sectors covering thus the whole $2\pi$ interval. This allows us to find the angle $\psi_n$ already in event plane, and provide the standard procedure to present $v_n$ as, in fact, the mean value of $v^{(s)}_n$ related to the "single" events. The presented results are based on averaging of hydrodynamic events for 200 such groups, thus having 10000 events overall. The results are represented for three centrality classes: $0-20\%$, $20-40\%$ and $40-60\%$.

 The total photon spectra for all centralities are shown in Fig. \ref{fig:spectra_all}. As it was discussed above, we compare the two scenarios: chemically equilibrated (iHKM) and chemically frozen (HKM). The resulting spectra show that there is an insignificant difference between two scenarios. The reason is that an appearance of additional chemical potentials of hadrons in a chemically frozen scenario changes the equation of state (EoS) compared to chemically equilibrated hydrodynamics, see Fig. 1 in Ref. \cite{original HKM} and Fig. 6 in Ref. \cite{HKM 2008}. This results in difference of the expansion rates at the different scenarios. As a result of the interplay between expansion rates at different EoS and chemical composition of the matter, the temperatures at the same energy densities are lower  in the case of chemically frozen expansion compared to the chemically equilibrated one, as it is demonstrated in Fig. 7 in Ref. \cite{HKM 2008}. Our detailed calculations demonstrate that this circumstance almost compensates effect of larger chemical potentials of hadrons in the photon spectra formation in chemically frozen scenario.

For the chemically equilibrium scenario the results without the hadronization emission are also demonstrated in Fig. \ref{fig:spectra_all}. The absolute value of HE contribution to the spectra is less for large centrality since the hadronization space-time layer is shrinking in more peripheral events.  As for the relative HE contribution it is only slightly changing with centrality.

In Fig. \ref{fig:spectrum_compare} the HE contribution is plotted along with the prompt and thermal (that includes also small thermal part from pre-thermal stage) contributions that forms the total photon spectra at $0 - 20 \%$ centrality. As one can see from Fig.\ref{fig:spectrum_compare},  in soft momentum region $p_T<3$ GeV$/c$ the HE contribution is the same as thermal photon radiation. So, one can conclude, like the authors of  Ref. \cite{low T},   that
the hadronization region is, probably, a key source of relatively soft  photon
emission.

\begin{center}
\begin{table}[ht]
\begin{tabular}{ |c|c|c|c|c| }
 \hline
   Centrality & Prompt yield & Thermal yield & Hadronization yield & Mean multiplicity \\
 \hline
 0-20\% & 0.2471 & 0.5 & 0.5499 & 567\\
 20-40\% & 0.0905 & 0.1727 & 0.2314 & 236\\
 40-60\% & 0.0265 & 0.0456 & 0.0751 & 98\\
 \hline
\end{tabular}
\caption{\small Photon yields for different sorts of photons and charged particle multiplicities for different centralities }
\label{tab:table_yields}
\end{table}
\end{center}

\begin{figure}
     \centering
     \includegraphics[width=1.0\textwidth]{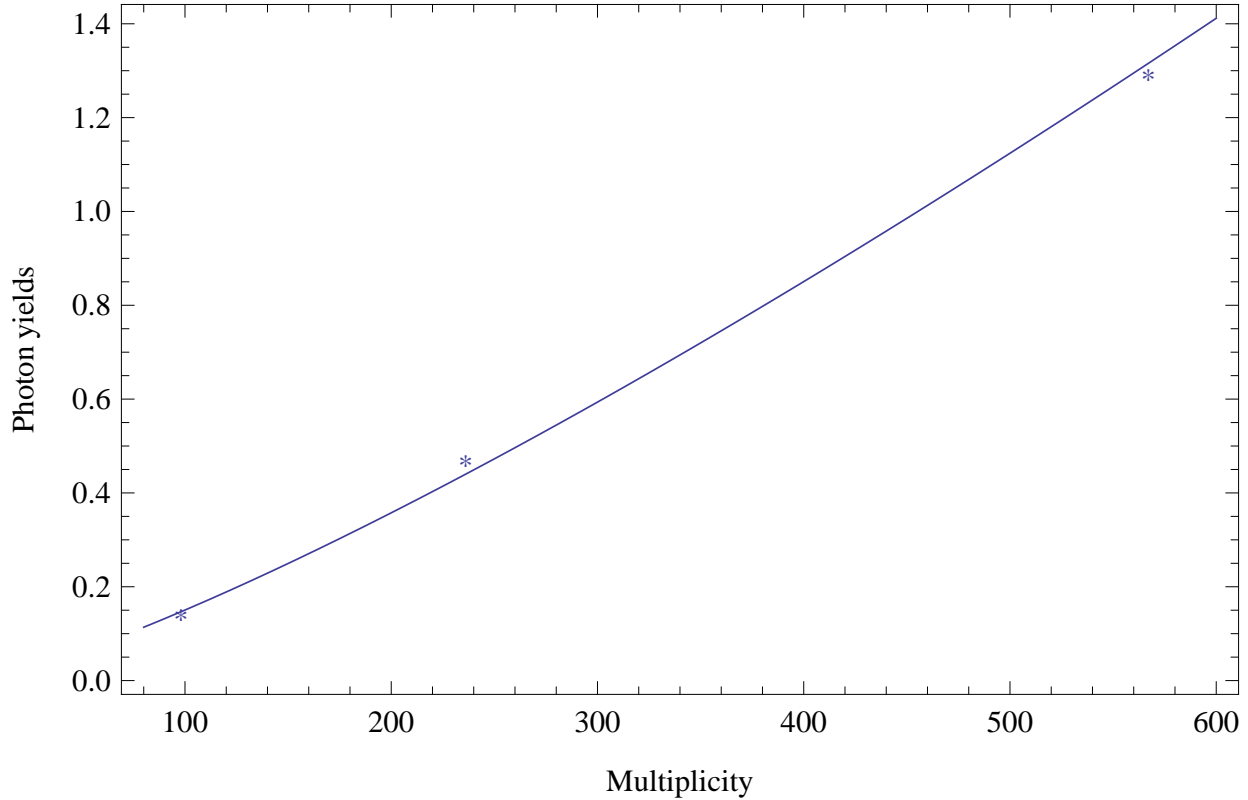}
     \caption{\small
A proportionality of the total direct photon yields (stars)  related to the 3 different centralities  in $Au+Au$ collisions at the top RHIC energy to $(\frac{dN_{ch}}{d\eta})^{1.25}$ (solid line).}
\label{fig:yields}
\end{figure}

In Table \ref{tab:table_yields} we demonstrate the dependence of the direct photons yields, related to the different  sources, on the charge particle multiplicity. It should be noted that our results are in agreement with recent observations made by the PHENIX Collaboration \cite{PHENIX-3}, namely, without details,   the total/soft  direct photon yield is proportional to charged particle multiplicity to the power 1.25. The comparison of our results with this PHENIX's finding is presented in Fig. \ref{fig:yields}.

\begin{figure}
     \centering
     \includegraphics[width=1.0\textwidth]{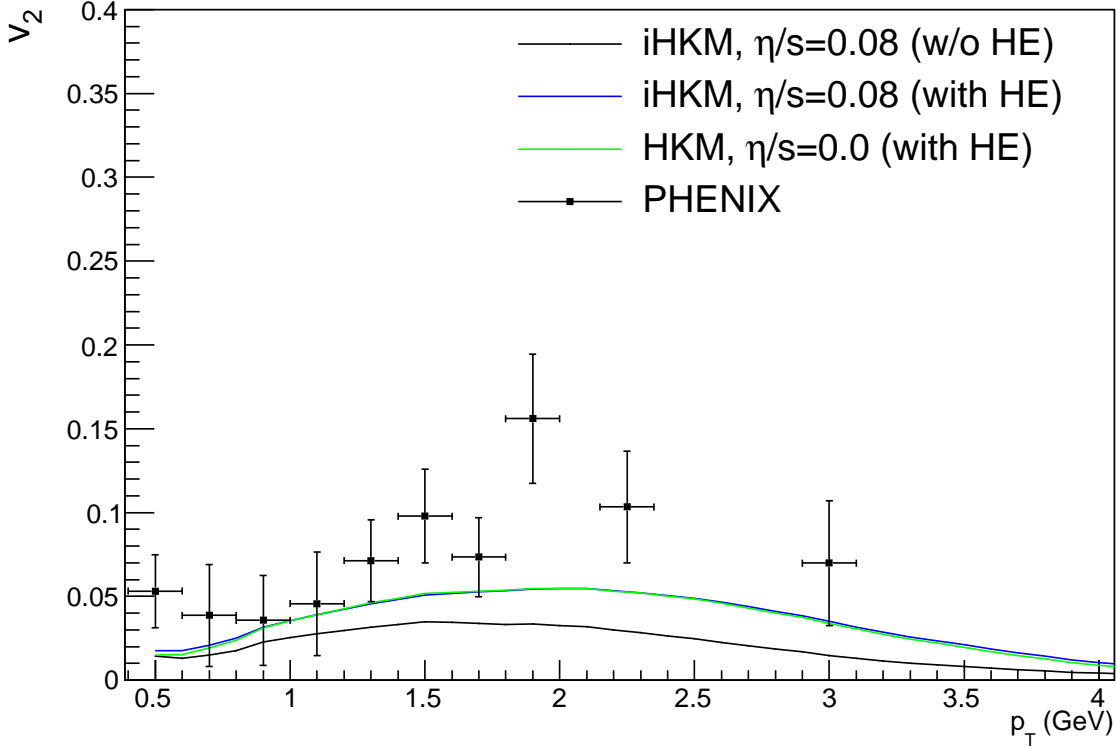}
     \caption{\small
Elliptic flow for 0-20\% centrality for the different models: iHKM chemically equilibrated with  contribution, iHKM chemically equilibrated without hadronization emission (HE) contribution, HKM chemically frozen at the hadron stage with continuous transition from hydrodynamics to hadron gas (with HE contribution). Experimental results are taken from \cite{RHIC-v2}.}
\label{fig:v2-0-20}
\end{figure}

\begin{figure}
     \centering
     \includegraphics[width=1.0\textwidth]{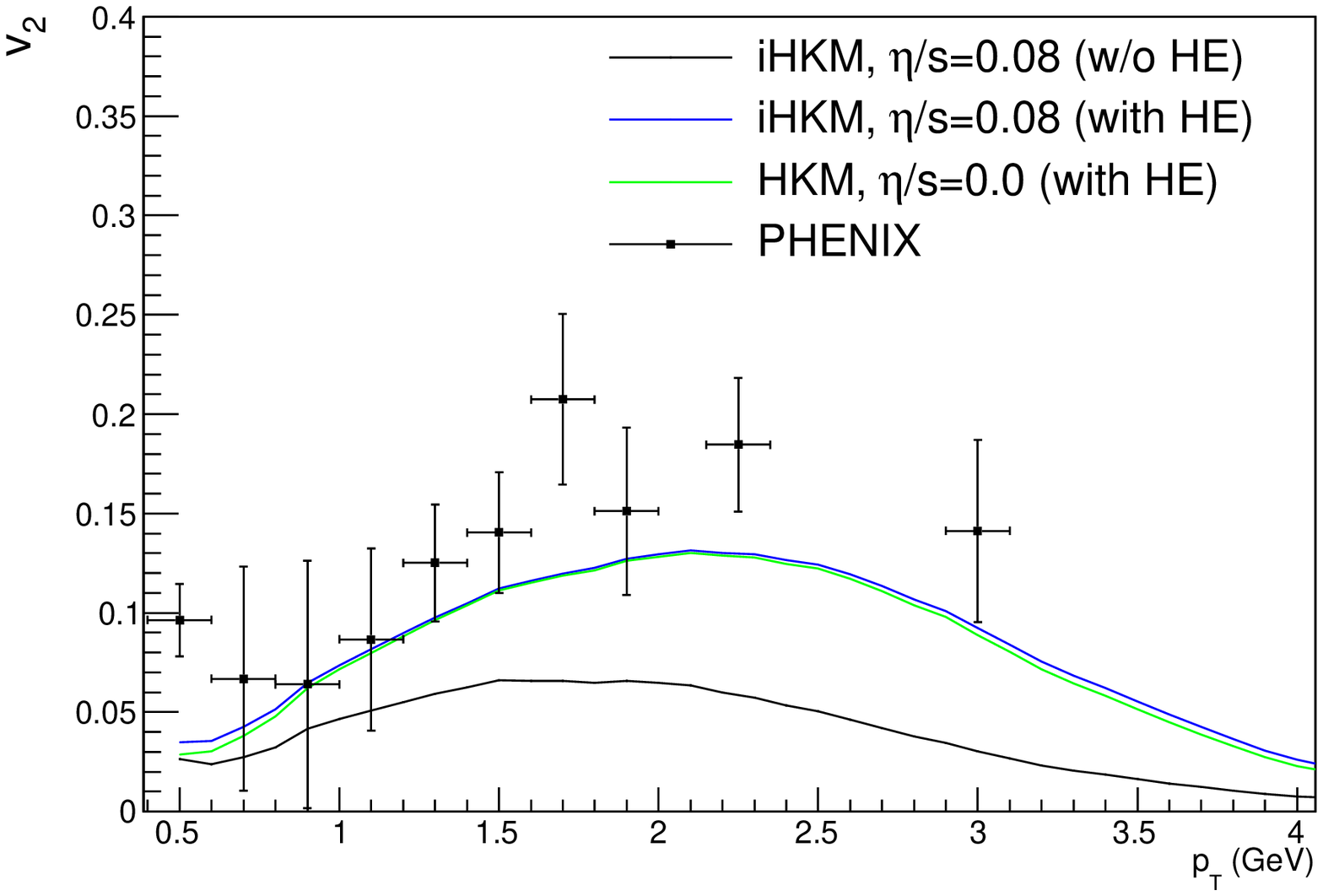}
     \caption{\small
Elliptic flow for 20-40\% centrality for the same conditions as in Fig. \ref{fig:v2-0-20}}
\label{fig:v2-20-40}
\end{figure}

\begin{figure}
     \centering
     \includegraphics[width=1.0\textwidth]{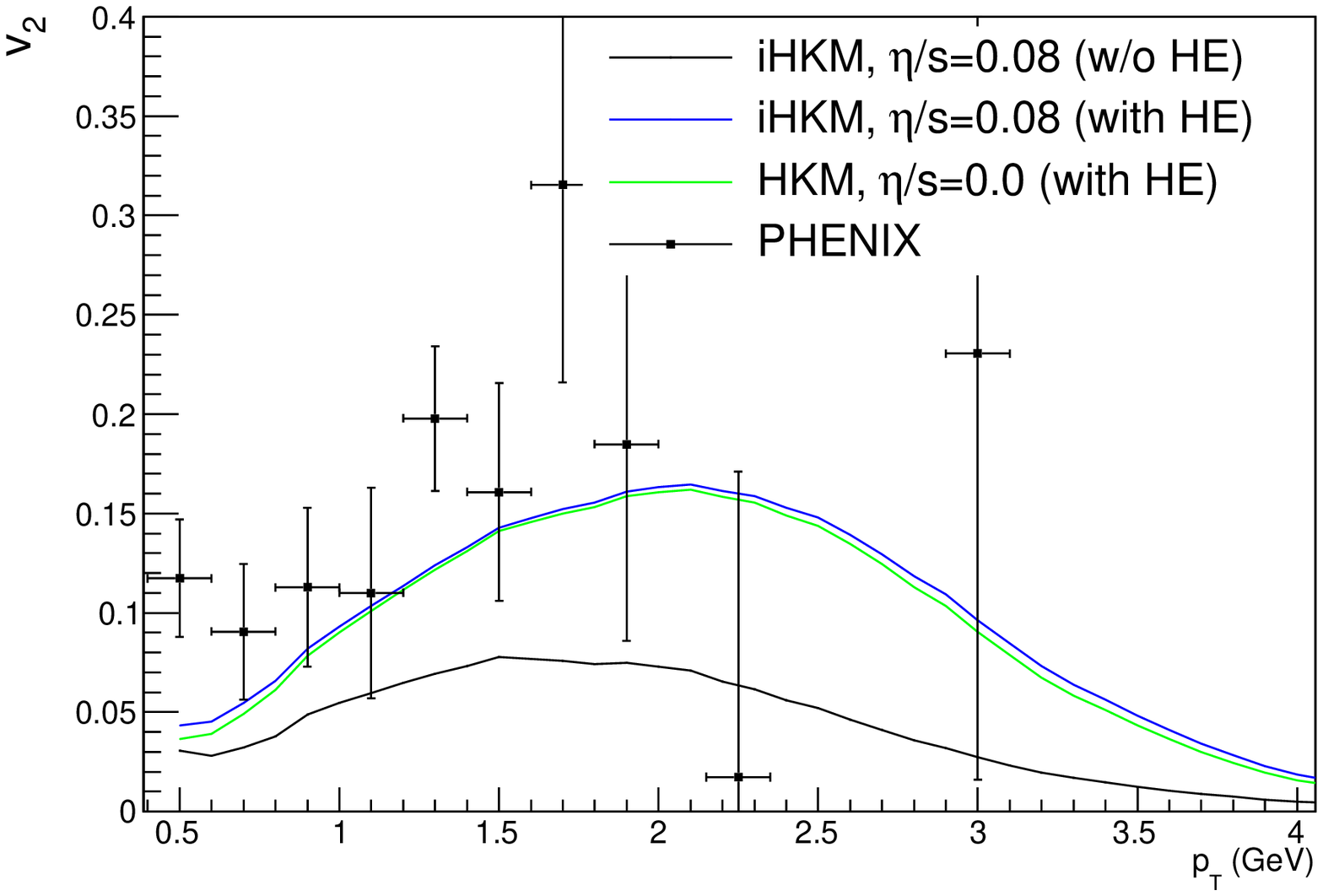}
     \caption{\small
Elliptic flow for 40-60\% centrality for the same conditions as in Fig. \ref{fig:v2-0-20}}
\label{fig:v2-40-60}
\end{figure}

  The photon elliptic flow dependence on photon momentum is shown on Figs. \ref{fig:v2-0-20} - \ref{fig:v2-40-60}. Two scenarios again give very similar descriptions, and HE emission gives a large impact on results in both scenarios.

\begin{figure}
     \centering
     \includegraphics[width=1.0\textwidth]{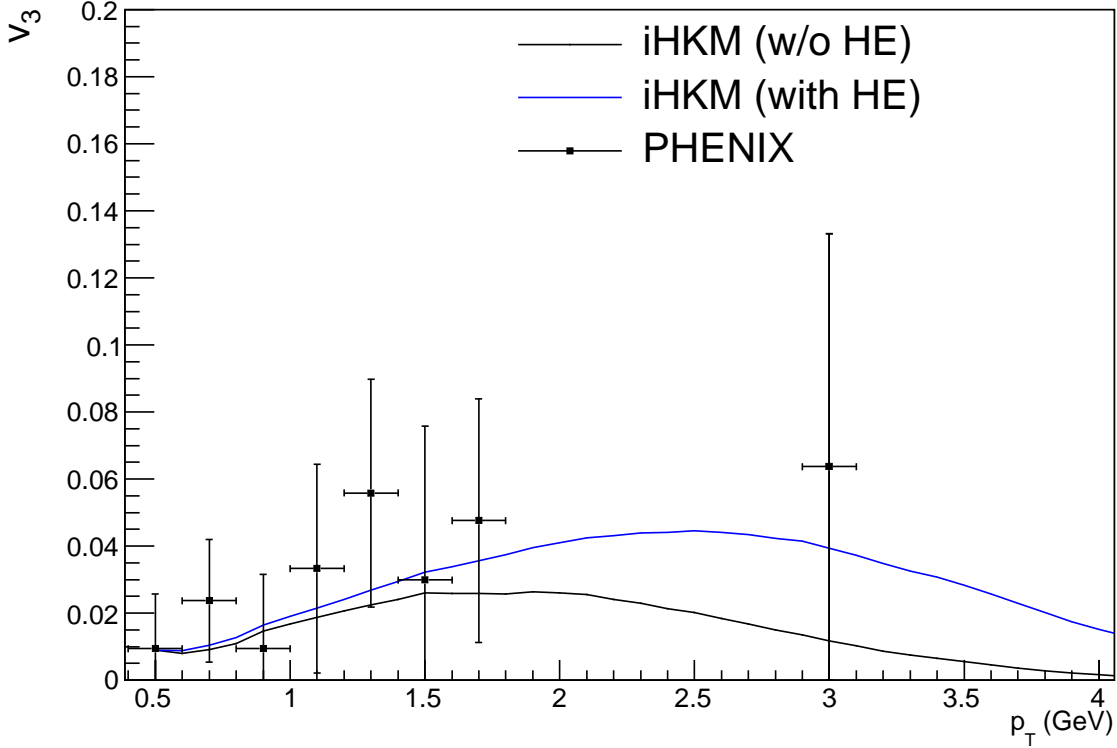}
     \caption{\small
Triangular flow for 0-20\% centrality for the different models: iHKM chemically equilibrated without hadronization emission (thermal and prompt photons only) contribution and iHKM chemically equilibrated with HE contribution. Experimental results are taken from \cite{RHIC-v2}.}
\label{fig:v3_0-20}
\end{figure}

\begin{figure}
     \centering
     \includegraphics[width=1.0\textwidth]{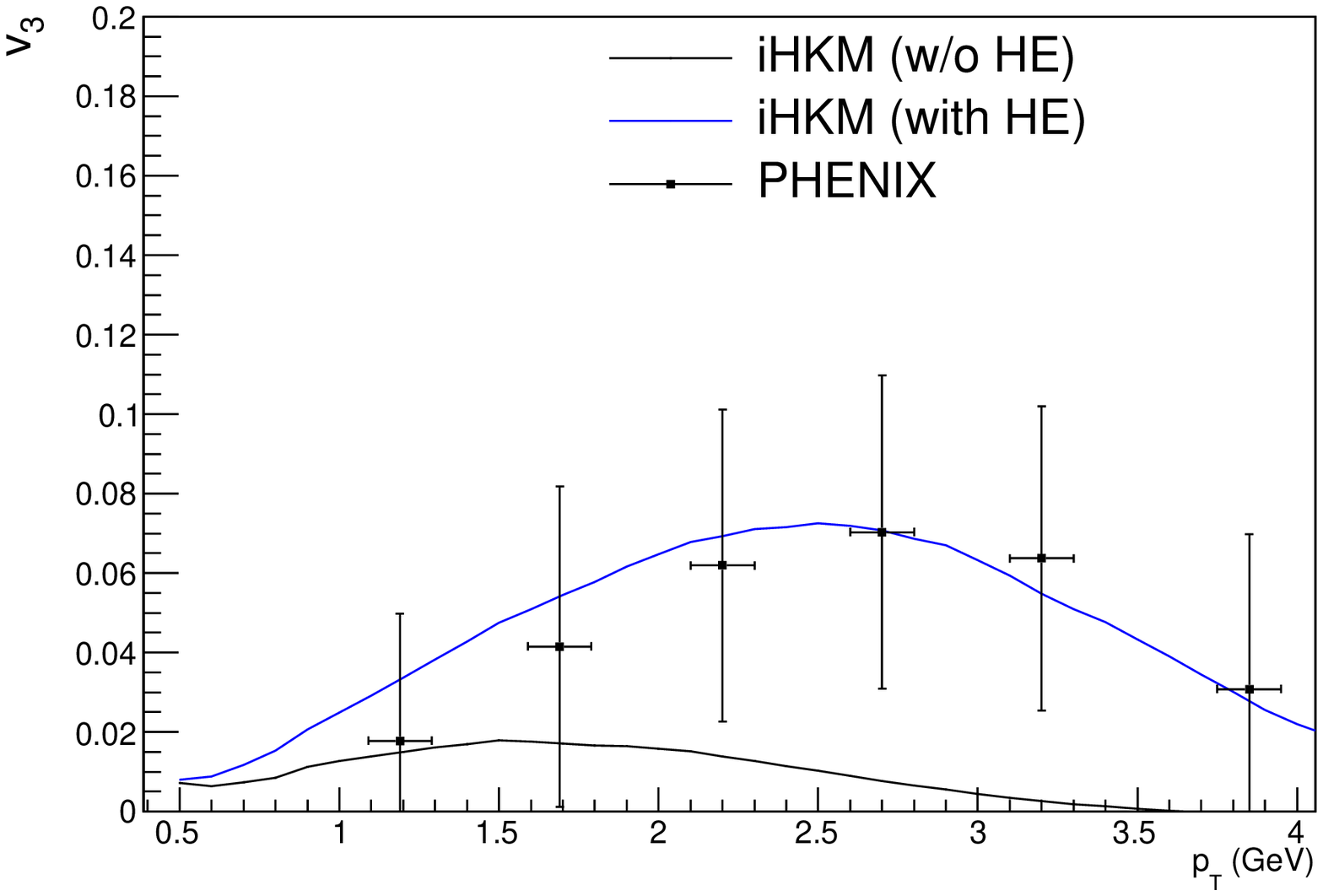}
     \caption{\small
Triangular flow for 20-40\% centrality for the same conditions as in Fig. \ref{fig:v3_0-20}}
\label{fig:v3_20-40}
\end{figure}

\begin{figure}
     \centering
     \includegraphics[width=1.0\textwidth]{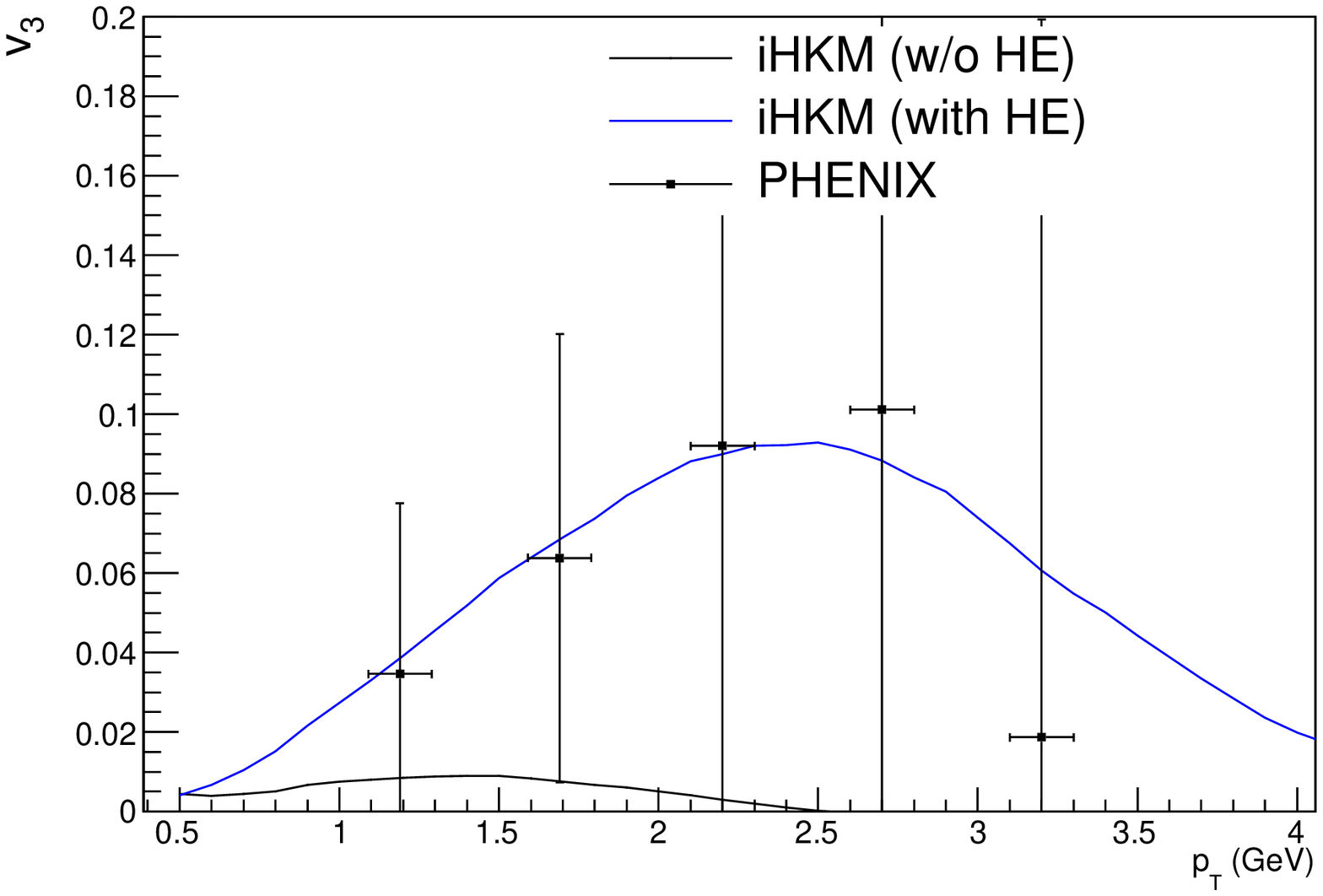}
     \caption{\small
Triangular flow for 40-60\% centrality for the same conditions as in Fig. \ref{fig:v3_0-20}}
\label{fig:v3_40-60}
\end{figure}

As for the triangular flow, presented at different centralities in Figs. \ref{fig:v3_0-20}	- \ref{fig:v3_40-60}, one can note that relative contribution of HE emission is growing strongly with centrality at the same fixed parameters of the model. The final results are in good agreement with experimental data  that is a pleasant surprise also for authors.

In general, our results, at least qualitatively,  are along with what already well understood in the electromagnetic probe community in the sense that increasing the rate of thermal photon production at low temperature would improve agreement with the direct photon spectra and $v_2$ data.  Such hypothetical enhancement of the thermal rate was investigated in Ref. \cite{low T}. The authors argued that an enhancement of photon rate by factor 2-3
especially in the pseudo-critical region leads to improving the data description for the spectra and elliptic flow. The model calculations are based on ideal hydro with initial transverse flow (free fitting function!) and ``sequential freezeout" with two quite different temperatures. All mentioned ingredients allows one to improve photon rate and spectra slope.  As for our consideration, it is based on much more developed evolutionary approach, with all parameters fixed from wide class of  hadronic observables at the top RHIC energy \cite{ihkm-rhic}. The iHKM contains all the stages of the nucleus collision process, has natural zero initial transverse velocity and continuous freeze-out. Being applied here to the photon business, an additional one-parametric source of photons is needed to describe photon spectra, elliptic and triangular flow at three centrality classes at RHIC. The effective parameter in this source for the photon rate $\beta=0.04$ is related to the specific processes of photon radiation that are connected to the confining interactions
at the hadronization transition and  the temporal width of the transition layer.
It is worthy noting, however, that our phenomenological parametrization of emission from this layer, could probably account, except radiation due to confinement, also for an additional photons due to  possible enhancement of the currently employed hadron emission rates in the pseudo-critical region, where the medium is expected to be most strongly coupled, as it was discussed in Ref. \cite{low T}.

\section{Summary}
\label{sec:summary}
In this paper we investigate the photon spectrum and its momentum anisotropy for heavy ion collisions at RHIC energy $\sqrt{s_{NN}}=200A$ GeV in the framework of the integrated hydrokinetic model. Incidentally, we treat different sources of photon emission aggregating, precisely, prompt photons, thermal photons radiating at the pre-thermal phase of evolution, thermal photons from quark gluon plasma and hadronic gas stages, and also we include the hypothetical hadronization emission (HE) of photons, emitting because of confinement process during expansion and cooling of a quark-gluon plasma. The last additional photon source  was introduced phenomenologically in the form of a Bose-Einstein distribution with only one effective parameter, which simultaneously takes into account the time-like width of the hadronization layer and the rate of HE. The  set of other iHKM parameters used here is just the same as that we have handled for successful description of hadronic bulk observables in $Au+Au$ $\sqrt{s_{NN}}=200$ GeV collisions at RHIC \cite{ihkm-rhic} including the particle yields and number ratios, pion, kaon, proton, antiproton spectra, transverse momentum anisotropy, and HBT-radii.

We compare the results of the two approaches to describe  the hadron gas emission supposing chemically equilibrated and chemically frozen evolution at the corresponding stage. Both approaches lead to quite similar results and they are mostly within the error bars for all centralities. The reason for this is in the compensatory mechanism: at the same energy densities which the system passes during the hadron stage of the evolution, the temperatures are higher at chemically equilibrated scenario compared with chemically frozen one, for which, however, the additional positive chemical potentials appear.

    It should be noted that our results are in agreement with recent observations made by the PHENIX Collaboration \cite{PHENIX-3}, namely: the total/soft yields are proportional to charged particle multiplicities. As it was demonstrated above,  iHKM  results are fully consistent with this observation. The reason is that the photon spectra at RHIC and LHC are well described (see Ref. \cite{ihkm-photon} and current results) with the same parameters for strongly interacting matter which brought a good agreement with data on charged particle multiplicities at different centralities as well as the other bulk hadron observables at the corresponding energies \cite{iHKM,ihkm-rhic, ihkm-lhc276, ihkm-lhc502, kaon-femto, Kstar}.

    The main result is that in order to describe photon spectra,  elliptic and triangular flow at different centralities within integrated HydroKinetic Model, describing well all bulk hadron observables, one needs an additional photon source that at fairly soft transverse momenta is compatible with "standard" thermal source. We associate this hypothetical source in the line that already discussed in the literature, namely, with "confinement photons". Of course, more detailed theoretical study of this mechanism is necessary. However, even preliminary it can be concluded that photon radiation in the pseudo-critical region is, probably, a key to solve the direct photon puzzle.

\section{Acknowledgments}
The research was carried out within NAS of Ukraine priority project "Fundamental properties of the matter in the relativistic collisions of nuclei and in the early Universe" (No. 0120U100935).  It is partially supported by Tomsk State University Competitiveness Improvement Program and  by NAS of Ukraine Targeted research program “Fundamental research on high-energy physics and nuclear physics(international cooperation)”.

\end{document}